   \documentclass[11pt,preprint2]{aastex}		

   \received{}
   \revised{}
   \accepted{}

   \slugcomment{Submitted to the Astronomical Journal.}

   \shorttitle{B3 0003$+$387: LSS at $z=1.47$?}
   \shortauthors{Thompson, Aftreth, \& Soifer}

   \normalsize

\begin{document}

\title{B3 0003$+$387: AGN Marked Large-Scale Structure \\ 
       at Redshift 1.47? \altaffilmark{1,2}}

\altaffiltext{1}{This paper uses data obtained at the WIYN Observatory, 
   a joint facility of the University of Wisconsin-Madison, Indiana University, 
   Yale University, and the National Optical Astronomy Observatories.}
\altaffiltext{2}{Based in part on observations obtained at the W. M. Keck 
   Observatory.}

\author{D. Thompson}
\affil{Palomar Observatory, California Institute of Technology\\
       MS 320-47, Pasadena, CA 91125}
\email{djt@mop.caltech.edu}

\author{O. Aftreth}
\affil{California Institute of Technology\\
       MS 175-48, Pasadena, CA 91125}

\and
\author{B. T. Soifer\altaffilmark{3}}
\affil{Palomar Observatory, California Institute of Technology\\
       MS 320-47, Pasadena, CA 91125}
\altaffiltext{3}{Also at the SIRTF Science Center, California Institute 
                 of Technology, Pasadena, CA 91125}

\begin{abstract}
We present evidence for a significant overdensity of red galaxies, as
much as a factor of 14 over comparable field samples, in the field of
the $z=1.47$ radio galaxy B3 0003$+$387.  The colors and luminosities
of the brightest red galaxies are consistent with their being at $z > 0.8$.  
The radio galaxy and one of the red galaxies are separated by 5\arcsec\ 
and show some evidence of a possible interaction.  However, the red 
galaxies do not show any strong clustering around the radio galaxy nor 
around any of the brighter red galaxies.  The data suggest that we are 
looking at a wall or sheet of galaxies, possibly associated with the 
radio galaxy at $z=1.47$.  Spectroscopic redshifts of these red galaxies 
will be necessary to confirm this large-scale structure.
\end{abstract}

\keywords{galaxies: clusters: general --- galaxies: distances and redshifts 
          --- large-scale structure of universe --- infrared: galaxies}


\section{Introduction}

As the most massive collapsed objects in the universe, clusters provide
a sensitive probe of the formation and evolution of structure in the
universe.  The presence of clusters or large-scale structure at and
beyond $z \approx 1$ can constrain scenarios for bottom-up structure
formation and fundamental cosmological parameters like $\Omega_{0}$ by
requiring cluster-level collapse earlier than was previously thought
possible.  Study of the individual galaxies in high-$z$ clusters can
also provide information on the cluster environment, merging and star
formation history in the cluster galaxies, and samples of cluster
ellipticals and spirals which could be compared with field galaxies at
the same redshift.

A number of confirmed or suspected clusters are known at redshifts 
above one, with some candidates extending up to $z \sim 3$ (see 
\citet{post_rev} for a recent review).  These include clusters around 
or foreground to high-$z$ active galactic nuclei 
\citep{MRC0316,Q0953,3C324,Hall&Green,Clements,B2_1335,Q1213}, a 
near-infrared selected cluster \citep{stanford}, a cluster associated with 
correlated quasar absorption line systems \citep{Fea96}, X-ray selected 
clusters \citep{Rosati,Has}, Lyman break galaxies \citep{steidel}, and 
a distant cluster which gravitationally lenses a background QSO 
\citep{Benitez}.

Identifying structures directly in visual wavelength data becomes
increasingly difficult at $z \gtrsim 1$.  The fainter galaxies and
strong cosmological K-corrections serve to reduce the contrast of the
high-$z$ galaxies against the dense background of faint blue field
galaxies at visual wavelengths.  The contrast at near-infrared
wavelengths, however, remains high even at $z \gtrsim 1$ because the
K-corrections are lower and well-behaved.  

In most high-redshift clusters, there is a clear concentration of the 
galaxies around what would be considered a classical cluster ``core,'' 
though \citet{Hall&Green} find some evidence for overdensities of red objects 
lacking a central concentration in fields of $z \sim 1.5$ radio-loud 
quasars.  The structures found by \citet{steidel} and \citet{Fea96} are 
extended over large regions of sky.  These fields may actually be sampling 
sheets or filaments of galaxies associated with large-scale structure, 
or perhaps a protocluster seen in a dynamically young state, before 
any significant virialization has occurred. 

In this paper we present optical and near-infrared observations of a 
high redshift radio galaxy field, covering a total area of 44.3 sq. 
arcmin, where an overdensity of red galaxies suggests the presence of 
some large-scale structure, perhaps a wall or sheet of galaxies 
rather than a collapsed cluster.  

\section{Observations and Reductions}

B3 0003$+$387 is a 1.36\,Jy radio source at 408\,MHz in the Third
Bologna Survey \citep{B3}.  The source is unresolved with 15\arcsec\  
resolution at 1.4\,GHz \citep{B3VLA}.  An optical spectrum obtained by
\citet{B3spec} indicates a redshift of 1.47, with strong emission lines
typical for powerful radio galaxies.  Preliminary images at optical and
near-infrared wavelengths of the B3 0003$+$387 field identified a
number of very red galaxies near the radio source, possibly indicating
the presence of a high redshift cluster, and prompting the deeper
observations presented in this paper.  All reductions were performed
using standard IRAF\footnote{IRAF is distributed by the National
Optical Astronomy Observatories, which are operated by the Association
of Universities for Research in  Astronomy, Inc., under cooperative
agreement with the National Science Foundation.} \citep{IRAF} commands.

\subsection{$K^\prime$-Band Imaging}

The $K^\prime$ data were obtained on UT 1997 June 20 using the Omega-Prime
camera \citep{OPrime} at the Calar Alto 3.5\,m telescope in Spain.
Omega-Prime  is a direct imaging camera with no cold pupil stop,
imaging onto a HAWAII 1024$^2$ HgCdTe array.  The image scale is
0\farcs396/pixel, giving an unvignetted field of view 6\farcm75
square.  We obtained 58 images in a random offset pattern designed to
maximize the pointing separation between temporally adjacent images.
Each image was the sum of ten 3-second integrations, giving a total
exposure time of 29 minutes.

We reduced the data using IRAF and a double-pass reduction algorithm.
After a first pass of sky-subtraction and flatfielding, residual 
gradients in the background were removed in the second pass using 
object masks to explicitly ignore pixels covered by known objects 
and bad pixels.  The reduced images were resampled to 
2048$^2$ pixels prior to integer pixel offsets and stacking, giving 
a final pixel scale very close to that of the WIYN data (0\farcs198 
vs. 0\farcs195).

\begin{figure}
   \plotone{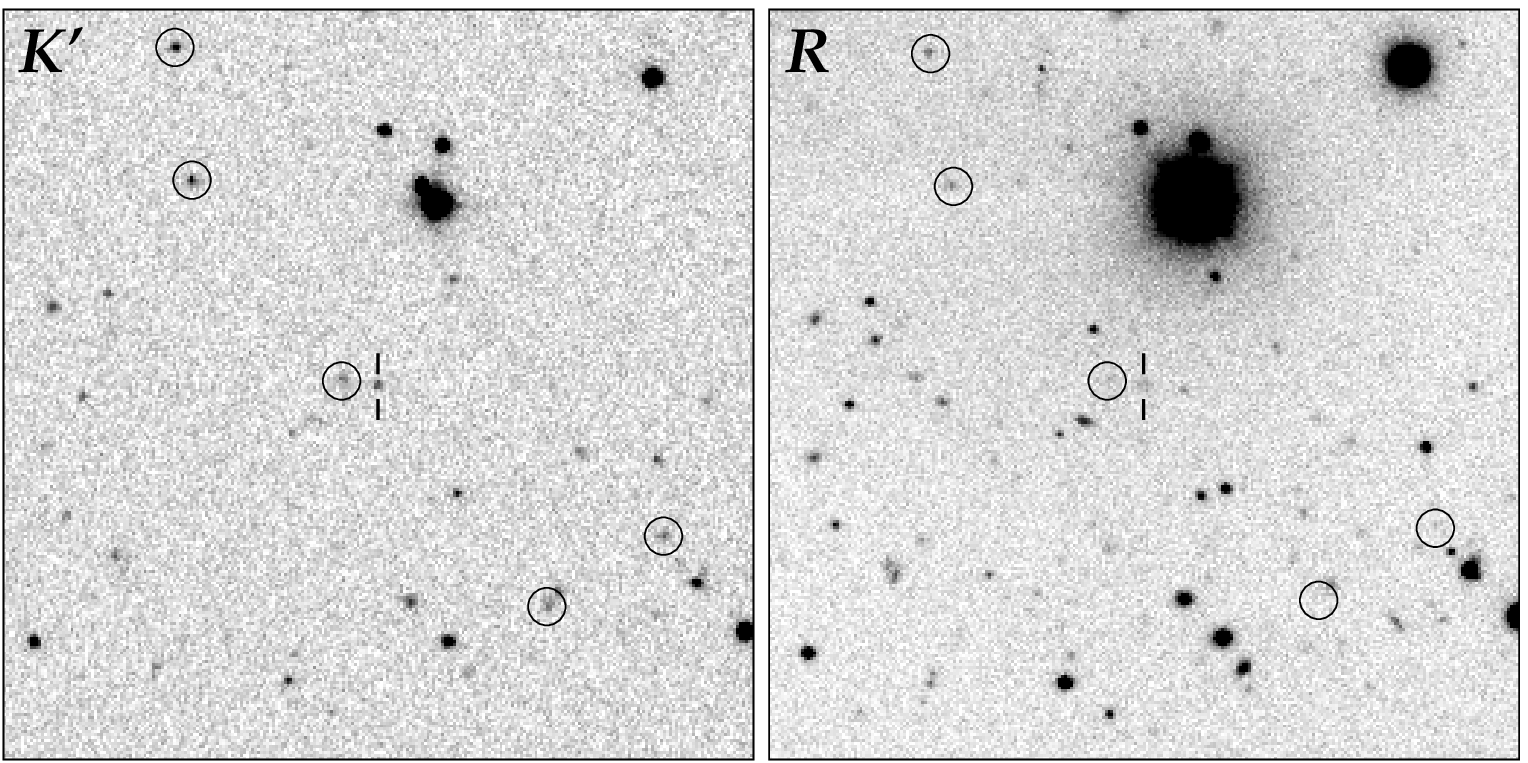}
   \caption{Two arcminute square subsections of the $K^\prime$ (left) and 
            $R$ (right) images, centered on B3 0003$+$387 (tick marks), 
            with north up and east to the left.  Several very red 
            galaxies with ($R-K^\prime$) $>$ 5 are circled.  
            \label{KpR2m}}
\end{figure}

The data were calibrated onto the Vega scale using several stars from
the UKIRT faint standards list \citep{UKIRTstds}.  The night was not
strictly photometric, but internal consistency of the standard star
measurements indicate that the systematic uncertainty in the
calibration is below 10\%.  The final $K^\prime$ image has a full-width
at half-maximum (FWHM) of 1\farcs1, and a 5-sigma point source
detection limit within an aperture diameter equal to twice the seeing
(2\farcs2) of $K^\prime = 20.31$.  A two arcminute square subsection of
the $K^\prime$ image, centered on the radio galaxy, is shown in the
left panel of Figure~\ref{KpR2m}.

\placefigure{KpR2m}

\subsection{$R$-Band Imaging}

Deep $R$-band data of the B3 0003$+$387 field were obtained through
service observing with the CCD imager at the 3.5m WIYN telescope on UT
1997 August 01 and UT 1997 August 05 in non-photometric but good seeing
conditions.  The imager uses a 2048$^2$ pixel CCD with an effective
plate scale of 0\farcs195 per 21\,\micron\ pixel, giving a 6\farcm66
square field of view which is well matched to the near-infrared image.
The data were taken through a Harris $R$ filter, which has a central
wavelength of 646\,nm and a FWHM of 153\,nm.  A series of short
exposures of the B3 0003$+$387 field and the \citet{Landoltstds}
standard fields SA92, SA110-L1 and Mark A were later obtained with the
same instrument under nearly photometric conditions on UT 1997 October
12 for calibration.

Standard CCD processing was used to remove the bias and flatfield 
each image.  Because the eight deep images were obtained on two 
different nights and under varying conditions, the signal-to-noise 
ratio (SNR) varies slightly from image to image.  We normalized the 
object photometry across all images, then derived weights
proportional to the variance in the rescaled sky background.  These eight
rescaled images were then combined with variance weighting, using 
cosmic ray and bad pixel masks.  

The calibration of the deep $R$ image was transferred from the 
shallower data taken on a nearly photometric night.  The standards 
indicate a stable night with $\sim$0\fm17 magnitudes of extinction 
over that expected under truly photometric conditions.  The final 
$R$-band image has a FWHM of 0\farcs8, and a 3-sigma point source 
detection limit within an aperture diameter equal to twice the seeing 
(1\farcs6) of $R_{Harris}=$25.58.  A two arcminute square subsection 
of the$R$ image, centered on the radio galaxy, is shown in the 
right panel of Figure~\ref{KpR2m}.

\subsection{Keck $K$-Band Imaging}

A deeper broadband $K$ image of the area around the radio galaxy was
obtained on UT 1997 August 20 using NIRC \citep{NIRC} on the Keck I
telescope.  NIRC reimages the infrared f/25 Keck focal plane onto a
$256^2$ Indium Antimonide detector at 0\farcs15 per pixel, for a
38\farcs4 square field of view.  We obtained a total of 29 1-minute
integrations, most in a regular 3\,$\times$\,3 pattern with
5\arcsec\ spacing.  Individual images consist of a sum of six 10-second
exposures.

\begin{figure}
   \epsscale{0.75}
   \plotone{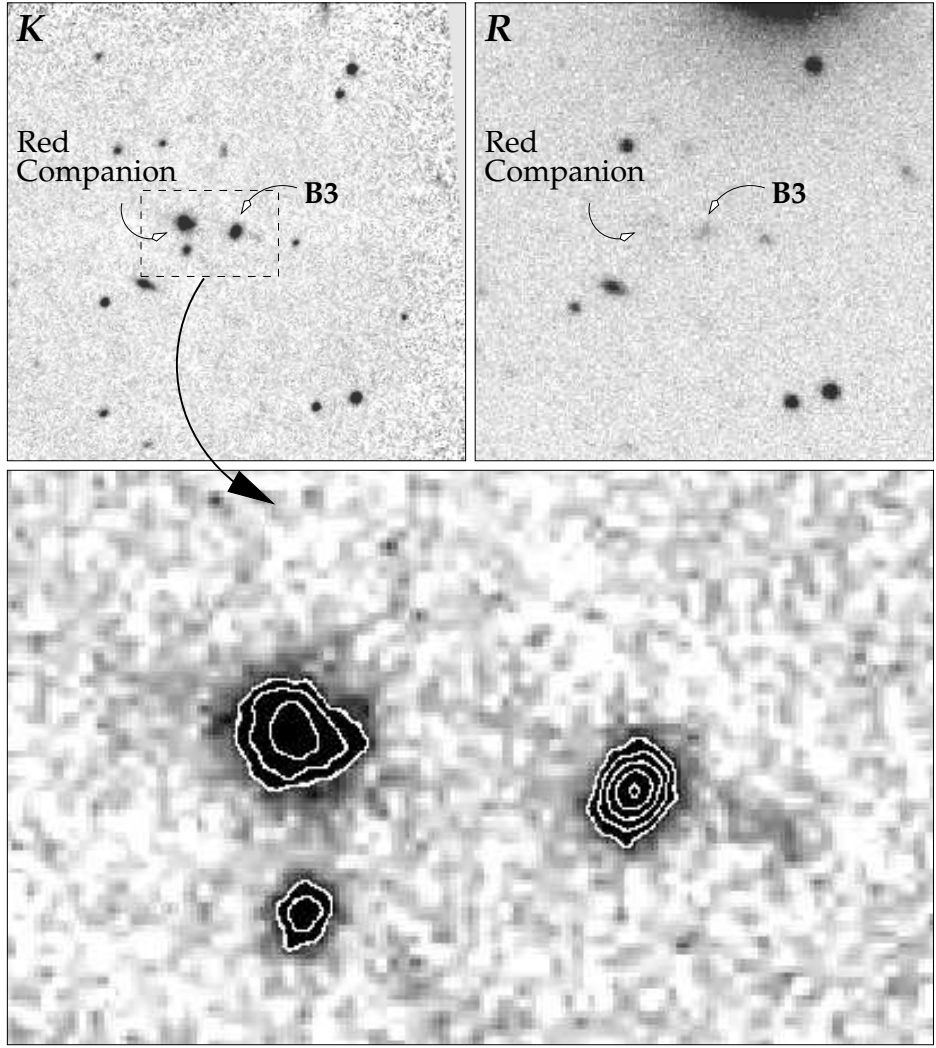}
   \caption{Comparison of the Keck $K$-Band (top left) and WIYN 
            $R$-band (top right) images in a 40\arcsec\,$\times$40\arcsec\ 
            area (1\% of the survey area) centered on B3 0003$+$387.  
            The radio galaxy is identified, as well as the brightest 
            ($K^\prime = 18.36$) and reddest 
            (($R-K^\prime$)\,$=$\,5\fm80) of 
            several very red objects evident in these images.  A 
            12\arcsec\,$\times$\,7\farcs5 subsection of the $K$-Band 
            image is also shown (bottom), with contours superimposed.  
            The red companion galaxy is extended in the direction of 
            the radio source, and additional faint features to the west of 
            the radio galaxy and to the east of the red companion are 
            visible in the grayscale image, suggesting that the pair of 
            galaxies may be interacting.
            \label{RKcomp}}
   \epsscale{1.00}
\end{figure}

Images were first sky-subtracted with a median stack of six to eight 
dithered images in the same field, then flatfielded with a skyflat 
derived from the same data.  The reduced images were combined with 
integer pixel offsets.  The data were calibrated onto the Vega scale 
using the \citet{IRstds} infrared standard stars.  The night was
relatively stable but not strictly photometric.  We estimate the 
systematic uncertainty in the zero point is again below 10\%.  
The $K$-band image has a FWHM of 0\farcs42, and a 5-sigma point-source
detection limit in the deep central part of the image of $K=22.44$.  The
$K$-band image and corresponding area of the $R$-band image are shown
in Figure~\ref{RKcomp}.

\placefigure{RKcomp}
    
\subsection{Astrometry}

Hubble Space Telescope Guide Star Catalog (GSC\footnote{GSC: Produced 
at the Space Telescope Science Institute under U.S. Government 
grant.}) stars around B3 0003$+$387 were used to derive fainter 
secondary position standards on a Digitized Sky Survey\footnote{DSS: 
Made by the California Institute of Technology with grants from the 
National Geographic Society, and produced at the Space Telescope 
Science Institute.} image of the field.  The absolute uncertainties 
in the astrometric solutions are $\Delta$\,$\approx$\,0\farcs3 in 
both the $R$ and $K^\prime$ images.  Objects in the Keck $K$-band image 
were matched to the $R$ and $K^\prime$ images by hand and were not fit 
with an astrometric solution.

The near-infrared object corresponding to B3 0003$+$387 was found to 
be at 00$^{h}$\,06$^{m}$\,20\fs71 $+$39$^\circ$\,00$^\prime$\,28\farcs1 
(J2000).  This agrees ($\Delta\alpha=0\farcs0,\ \Delta\delta=1\farcs1$) 
to well within the positional uncertainties from the VLA in 
C-configuration at 1.4\,GHz (FWHM\,$=$\,15\arcsec), used for the 
B3VLA Survey \citep{B3VLA}.  This is also the object for which 
\citet{B3spec} obtained the $z = 1.47$ redshift.  We therefore 
consider this identification of the radio galaxy secure.

\subsection{Object Catalog}

We created a catalog of all objects detected in the $K^\prime$ image, 
and used the APPHOT package in IRAF to extract the magnitudes 
within a 4\arcsec\ diameter aperture.  In a small number of cases, 
the photometry for close pairs or multiples was extracted by hand 
after subtracting out the contaminating objects.  The APPHOT 
positions in the $K^\prime$ image were transformed into coordinates on 
the $R$ image using our astrometric solution, and the 
corresponding $R$-band magnitudes extracted.  Any $K^\prime$-selected 
objects that were not detected in $R$ were assigned the three-sigma 
limit of $R=25.58$.  Because the $R$ image is much deeper than 
the $K^\prime$ image, there are large numbers of faint, relatively blue 
objects detected in the $R$ image which do not have a 
corresponding detection in the $K^\prime$ and therefore do not appear 
in the object catalog.  

The final catalog (Table~\ref{catalog}) contains the $K^\prime$ and $R$ 
centroids, magnitudes, uncertainties on the magnitudes (statistical 
only), and J2000 coordinates for all $K^\prime$-selected objects covering 
an area of 44.3 square arcminutes, or 1.23$\times10^{-2}$ square 
degrees.  There are 421 objects in the catalog, although 4 were 
deleted because of unrecoverable contamination in the $R$ image.  
The catalog and survey images are available through the electronic 
version of the Astronomical Journal.

\placetable{catalog}

\section{Discussion}

In the absence of spectroscopic redshifts for these red galaxies, we 
must rely on the imaging and photometry in hand to determine if 
there is a cluster or some sort of large-scale structure present in this 
field.  Specifically, we address the overdensity of red objects in the 
field relative to a field sample, the spatial distribution of these red 
objects, and what their redshift might be.  

We use the data from two widely-separated fields obtained as part of the 
Calar Alto Deep Imaging Survey (CADIS: see \citet{CADIS} for a description 
of the survey) as a control sample for comparison to the B3 data.  The CADIS 
fields were chosen to be ``blank'' fields at high galactic latitude, free 
of bright stars or galaxies, as well as known galaxy clusters and therefore 
should be representative of the field population of red objects.  Each 
field is $\sim$3 times the area of our data, and the combined area 
covers 6.4 times the area of the B3 0003$+$387 field.  The CADIS and 
B3 infrared data were obtained with the same instrument.  Both datasets 
use non-standard $R$-band filters, but the differences are negligible and 
the two data sets are directly comparable.  \citet{CADIS_EROs} presented 
initial results on the surface density of field extremely red galaxies 
from the CADIS data. 

For the following discussion, we assume that all of the red objects 
in the radio galaxy field are galaxies.  Most ($\sim$75\%) are 
resolved in the infrared image, but contamination of the sample by 
low mass stars, from late M dwarfs through L and T type brown dwarfs, 
is still possible.  Our survey field has a relatively high galactic 
latitude ($l_{II} = 113^\circ, b_{II} = -23^\circ$) and only covers 
a volume of 1.25\,pc$^3$ to a distance of 100\,pc.  We expect much less 
than one low mass star within this volume, given recent determinations 
of their space density \citep{Reid99,Herbst99}.

\subsection{An Excess of Red Galaxies}

\begin{figure}
   \plotone{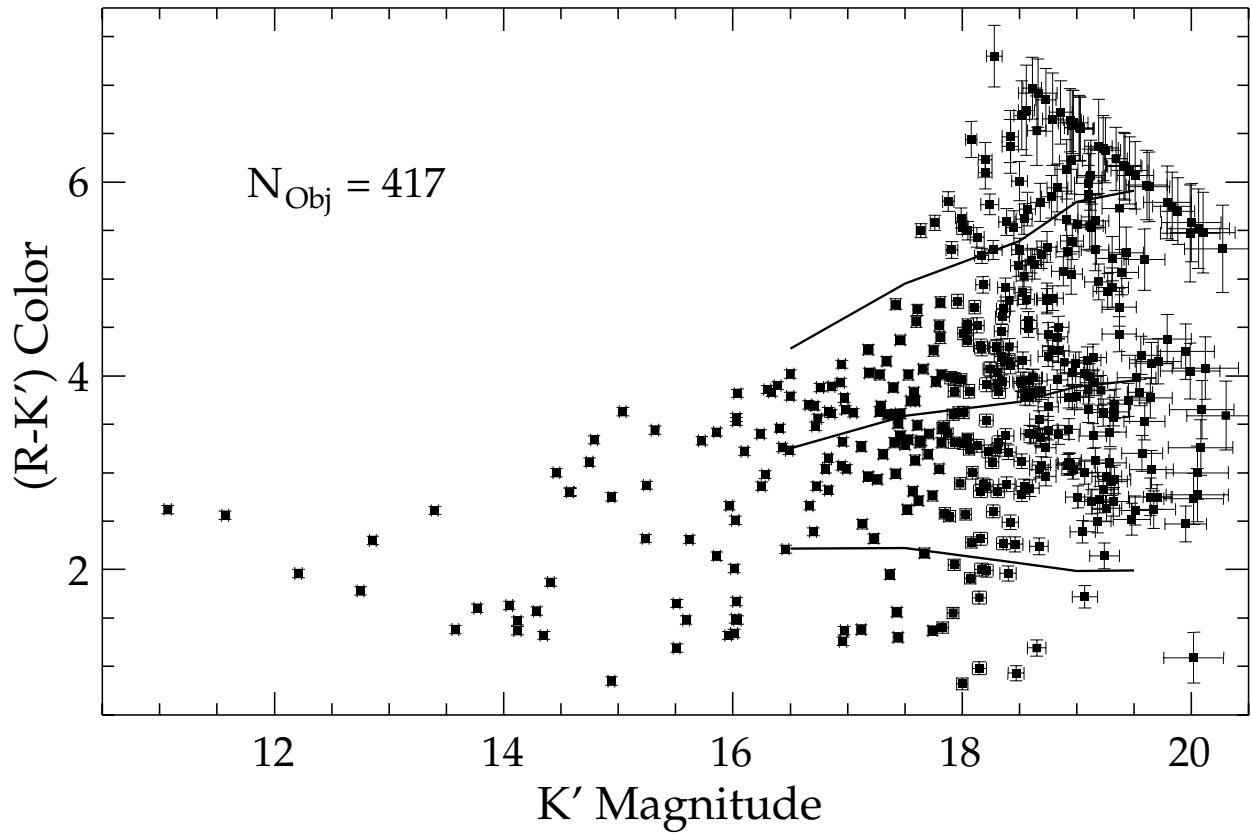}
   \caption{The full color-magnitude diagram for objects in the 
            B3 0003$+$387 field, with error bars showing the 
            $\pm1\sigma$ uncertainties in both the $K^\prime$ magnitude and 
            ($R-K^\prime$) color.  The three solid lines reproduce the median 
            and $\pm\,1.75\,\sigma$ distribution of colors from 
            \citet{CADIS_EROs}.
            \label{cmd1}}
\end{figure}

A color-magnitude diagram for the B3 0003$+$387 field is shown in
Figure~\ref{cmd1}.  We also plot the median and $\pm\,1.75\,\sigma$
distribution of colors from \citet{CADIS_EROs}, derived from the CADIS
data.  Several features in Figure~\ref{cmd1} are immediately obvious.
First, there are a number of objects with blue colors ($R-K^\prime <
2^{\rm m}$) which span the full range of $K^\prime$ magnitudes.  These
are typically galactic subdwarf stars.  Second, no objects brighter
than $K^\prime = 17.5$ appear {\em redder} than the $+\,1.75\,\sigma$
line.  However, this is consistent with purely statistical
fluctuations, as only 4 objects would be expected to lie at
$>\,1.75\,\sigma$ in our sample of 108 with $K^\prime < 17.5$.
Finally, there are a large number of objects with  $K^\prime > 17.5$
and redder than the $+\,1.75\,\sigma$ line.

\placefigure{cmd1}

\begin{figure}
   \plotone{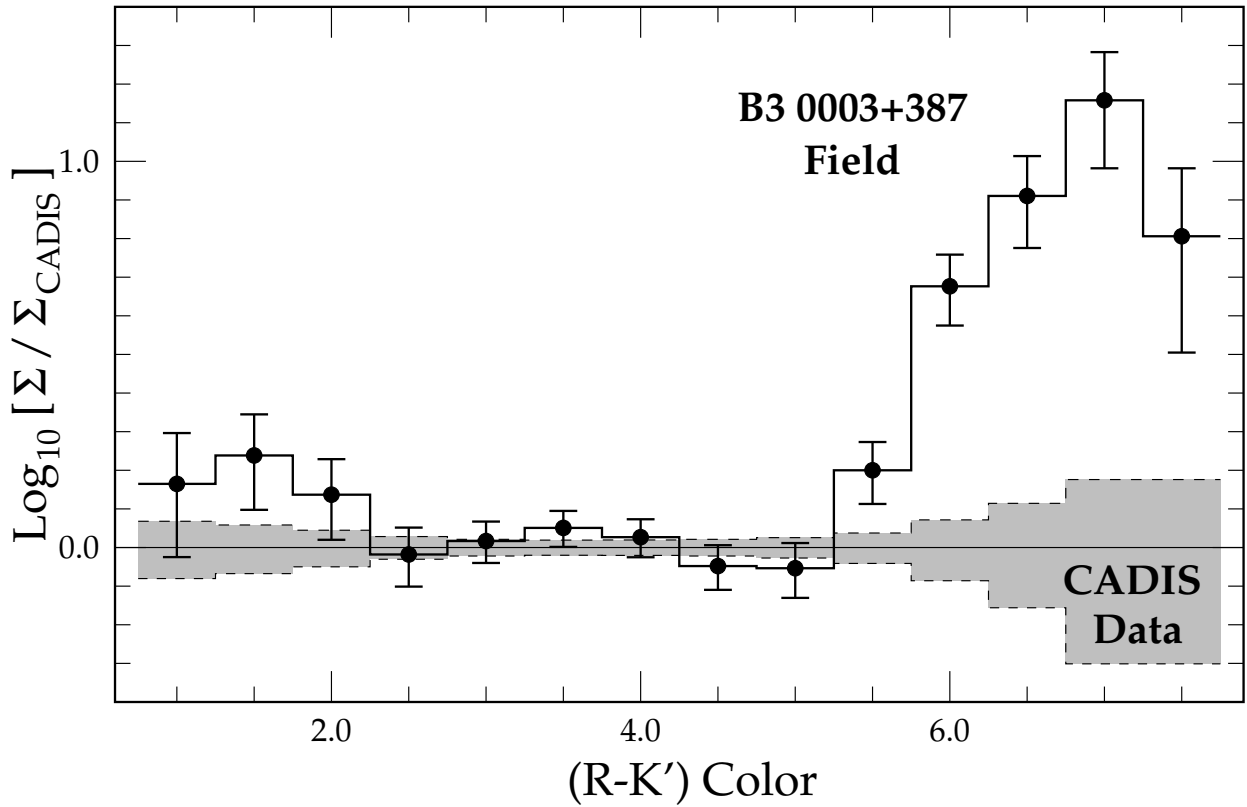}
   \caption{The logarithm of the normalized surface density of galaxies 
            in the magnitude range $17.0 \leq K^\prime \leq 19.0$, 
            plotted as a function of the ($R-K^\prime$) color.  The data were 
            normalized by an equivalent set of data from CADIS covering 
            over six times the area of the B3 field.  The gray region 
            represents the 1$\sigma$ Poisson uncertainties on the 
            self-normalized CADIS data.  Adjacent bins are {\em not}
            statistically independent due to the sampling method (see 
            text).  A significant overdensity of red galaxies, up to a 
            factor of 14, is evident in the data.  
            \label{NSDens}}
\end{figure}

Is this excess of red galaxies in the B3 0003$+$387 field significant?
We compared the surface density of galaxies in the B3 field with the
combined control fields from CADIS to determine the overdensity.  We
first selected galaxies from both sets of data in the $17.0 \leq K'
\leq 19.0$ magnitude range.  This covers the relevant range for the
red galaxies while avoiding problems with incompleteness at fainter
magnitudes.  We then sampled the data in 1 magnitude bins of color,
spaced every half magnitude.  Adjacent bins therefore are {\em not}
statistically independent.  We show in Figure~\ref{NSDens} the
logarithm of the surface density of galaxies in the two sets of data as
a function of the ($R-K^\prime$) color, {\em normalized} by the CADIS data.
The 1$\sigma$ uncertainties from Poisson counting statistics are also
indicated.

\placefigure{NSDens}

While the surface densities in the two fields agree to within the 
uncertainties for ($R-K^\prime$)\,$\leq$\,5$^{\rm m}$, there is a significant 
excess of galaxies at redder colors.  The most significant bin, with a 
3.7$\sigma$ excess, is at ($R-K^\prime$)\,$=$\,6$^{\rm m}$ (covering the 
range 5\fm5\,$\leq$($R-K^\prime$)\,$\leq$6\fm5).  The highest overdensity 
of red galaxies occurs at ($R-K^\prime$)\,$=$\,7$^{\rm m}$.  Combining the 
overdensities in statistically independent bins shows a $\sim 4.5 \sigma$ 
excess for ($R-K^\prime$)\,$\geq$\,5\fm5.  This suggests that this B3 field 
is not representative of the general field population, and that there is a 
large overdensity of red galaxies in this field. 

This overdensity of red galaxies may indicate the presence of a high 
redshift cluster, but there is no clear red sequence \citep{redseq1,redseq2} 
of cluster ellipticals visible in the color-magnitude diagram 
(Figure~\ref{cmd1}).  We next consider the spatial distribution of these 
red galaxies in the B3 field.  
 
\subsection{Red Galaxy Spatial Distribution}

Our survey field covers 6\farcm46\,$\times$6\farcm86.  For an Einstein-de
Sitter Universe ($\Omega_M = 1$, $\Omega_\Lambda = 0$; with $H_0 = 
100 h$\,km\,s$^{-1}$\,Mpc$^{-1}$) at the redshift of the radio galaxy, 
this translates to field of 1.66$h^{-1}$\,Mpc $\times$ 1.76$h^{-1}$\,Mpc, 
considerably larger than a typical galaxy cluster core.  
If an evolved, relaxed cluster were in our field, we would expect to 
see a significant concentration of galaxies around the cluster core.

\begin{figure}
   \plotone{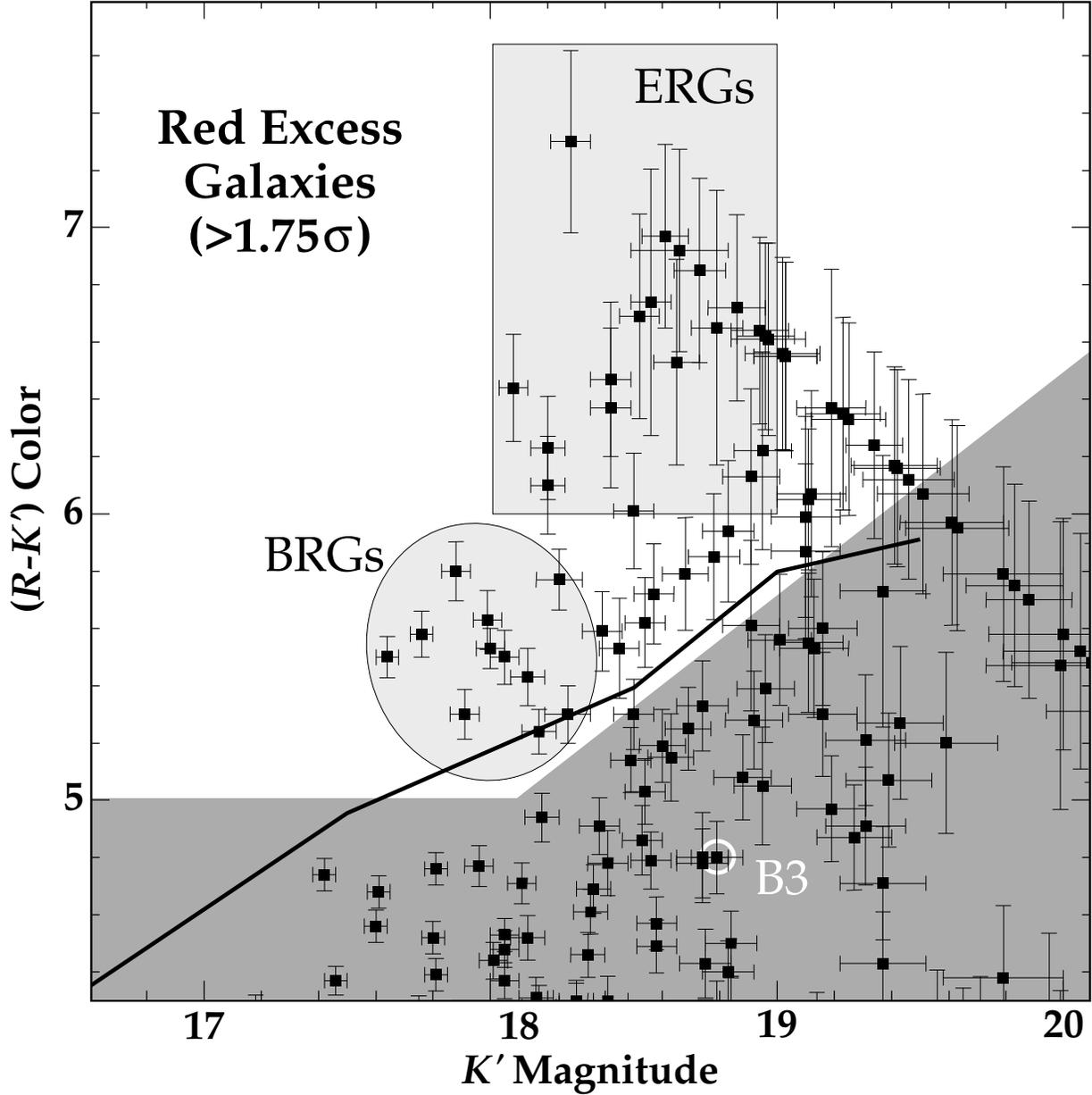}
   \caption{Detail from Figure~\ref{cmd1}.  The solid
            line is the $+\,1.75\,\sigma$ line from \citet{CADIS_EROs}.  
            All objects above the dark shaded region 
            (($R-K^\prime$)\,$=$\,0.735$K^\prime$\,-\,8.23 for 
            $K^\prime > 18^{\rm m}$) 
            are the red excess galaxies (REGs).  The light shaded oval 
            and rectangle indicate the brightest red galaxies (BRGs) and 
            extremely red galaxies (ERGs), subsets of the REGs.  The 
            radio galaxy (B3) is also indicated.  
            \label{cmd2}}
\end{figure}

For purposes of the discussion in this section, it is useful to divide 
up the group of red galaxies into three subsets: the brightest 
red galaxies, the extremely red objects, and the red excess galaxies.  
We show these subsets graphically in Figure~\ref{cmd2}, which reproduces 
the faint red end of the color-magnitude diagram (Figure~\ref{cmd1}) 
in greater detail.  

\placefigure{cmd2}

The brightest red galaxies (BRGs) subset consists of the 11 galaxies
with 17.5 $\leq$ $K^\prime$ $\leq$ 18.3 and 5\fm2 $\leq$ ($R-K^\prime$)  
$\leq$ 5\fm8 in the B3 0003$+$387 field.  Eight of these are resolved in 
the infrared image.  Such a group of objects with similar magnitudes 
and colors suggests that they may have a similar evolutionary history, 
possibly associated in a cluster.  In addition, their redder than normal 
colors imply that these BRGs are likely to be at high redshift.

The extremely red galaxies (ERGs) subset includes 20 galaxies with
($R-K^\prime$) $\geq$ 6\fm0 and $K^\prime \leq 19.0$, adopting 
the definition from \citet{CADIS_EROs}.  Fifteen of these ERGs are 
resolved.  The B3 0003$+$387 field covers 44.3 square arcminutes of 
sky, giving a local surface density for the ERGs of 
$0.45\pm0.10$\,arcmin$^{-2}$, a factor of $\sim$12 higher than that 
found in the CADIS data \citep{CADIS_EROs}.

The red excess galaxies (REGs) subset is defined as the galaxies with
an ($R-K^\prime$) color greater than $+1.75\sigma$, where the mean color and
uncertainty are taken from \citet{CADIS_EROs}.  This sample includes
both the BRG and ERG subsets.  For ease of use, we approximate the
$+1.75\sigma$ limit with the relation: ($R-K^\prime$) $\geq$ 0.735\,K$^\prime$
$-$ 8.23 for $K^\prime > 18^{\rm m}$, and ($R-K^\prime$) $\geq$ 5 for
$K^\prime \leq 18^{\rm m}$.

\begin{figure}
   \plotone{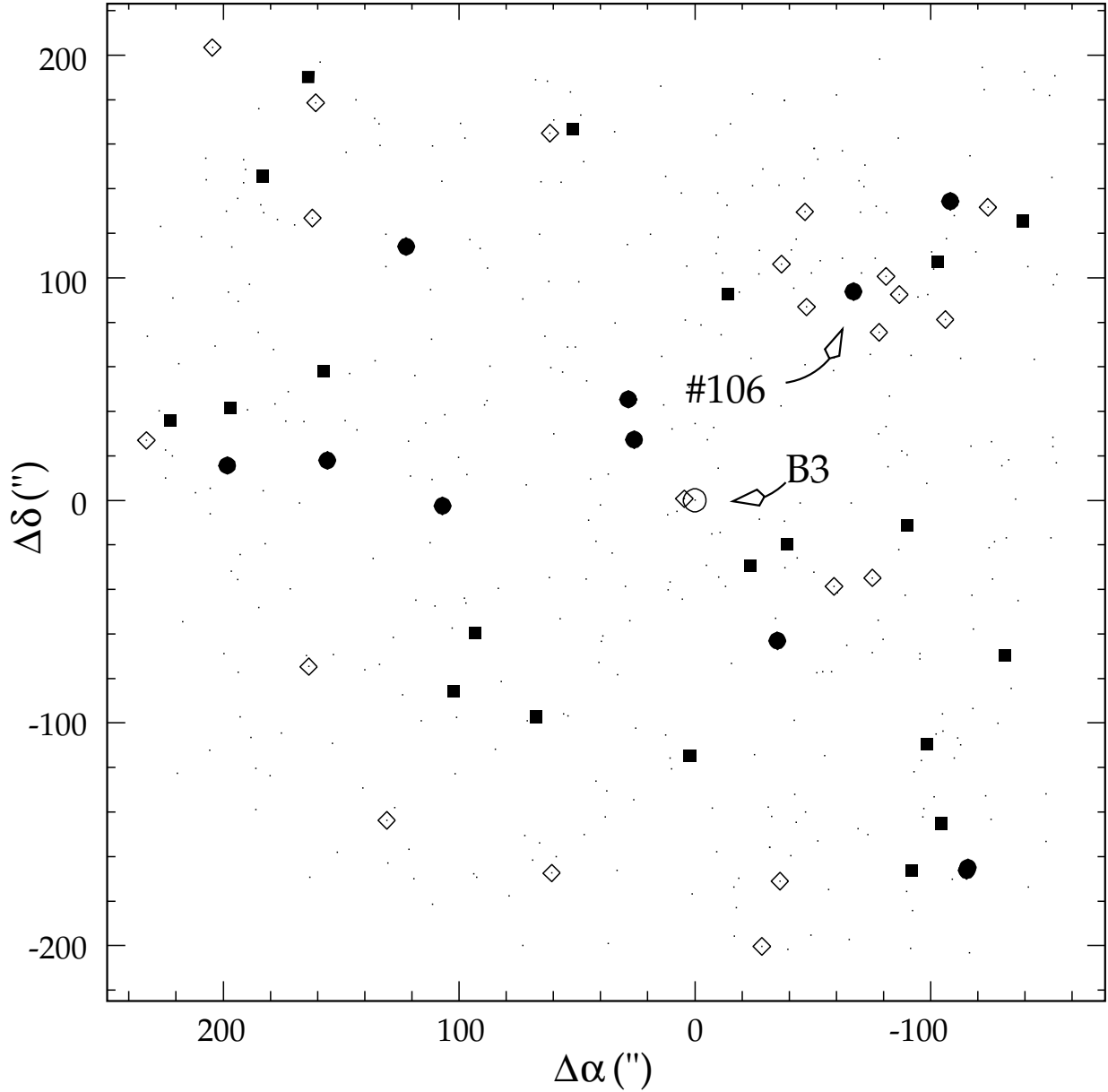}
   \caption{The positions of all objects (points) relative to 
            B3 0003$+$387.  Also indicated are the radio galaxy 
            (circled), the BRGs (filled circles), the ERGs (filled 
            squares), and all remaining red objects with 
            ($R-K^\prime$)\,$>$\,1.75\,$\sigma$ (open diamonds).
            \label{B3_pos}}
\end{figure}

The spatial distribution of these subsets is shown in Figure~\ref{B3_pos}, 
where we plot the positions of all objects in the B3 0003$+$387 field, 
highlighting the three subsets defined above.  What is immediately obvious 
in this plot is that there is no strong concentration of red galaxies 
around B3 0003$+$387.  This is in marked contrast to other high redshift 
clusters studied in the near infrared \citep{stanford,Rosati}.  We note 
that the infrared data around those clusters is considerably deeper 
(1\fm5 -- 2\fm0) than here and thus would be able to select cluster 
galaxies significantly further down the luminosity function.  
Our data do not provide evidence that B3 0003$+$387 lies in the core 
of a high-redshift cluster.

\placefigure{B3_pos}

There is a possible concentration of red excess galaxies around 
one of the BRGs, catalog number 106, located to the northwest of the 
radio galaxy.  This galaxy, with $K^\prime = 18.17$ and 
($R-K^\prime$) $=$ 5\fm24, is not exceptional in our sample: it is the 
bluest and one of the fainter galaxies in the BRG subset.  There are 
eight objects from the REG subset, one of which is an ERG, within a 
circular aperture of 45$^{\prime\prime}$ radius.  This is a factor of 
$\sim$3 less than in the \citet{stanford} and \citet{Rosati} clusters, 
assuming all of the red excess galaxies are associated.  

\begin{figure}
   \epsscale{0.75}
  \plotone{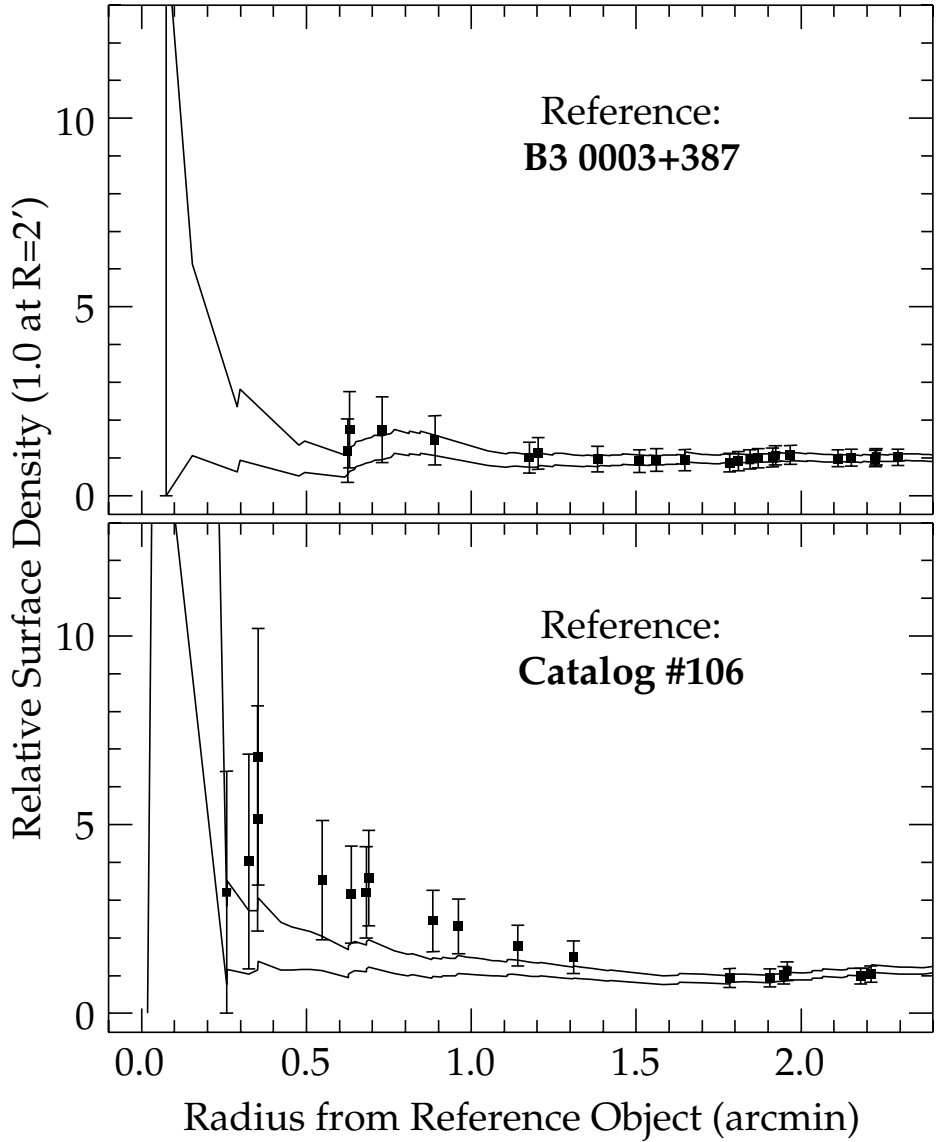}
   \caption{The radial surface density of red excess galaxies (filled 
            circles) around B3 0003$+$387 (top panel) and galaxy \#106 
            (bottom panel).  The solid lines plot the $\pm1\sigma$ 
            surface density of all field objects with $17.0 \leq 
            K^\prime \leq 19.5$, centered on the radio galaxy or 
            \#106.  The different datasets have been normalized to 
            unity at r $=$ 2$^\prime$. 
            \label{RSD}}
   \epsscale{1.00}
\end{figure}

In order to place more quantitative limits on the concentration of 
galaxies around any galaxies in our field, we investigated the
radial surface density distribution of red galaxies with respect to
various reference objects (BRGs and ERGs).  In no case was there a
statistically significant overdensity.  Figure~\ref{RSD} shows the
relative radial surface density of the red excess galaxies around 
B3 0003$+$387 and galaxy \#106.  In both cases, the data are normalized 
to unity at a radius of two arcminutes.  The distribution of red excess 
galaxies around B3 0003$+$387 is completely consistent with the bluer 
field galaxies selected in the same magnitude range.  While galaxy 
\#106 does show an overdensity of a factor of six to eight, it is not 
statistically significant (less than 2$\sigma$).  Note that both galaxies 
have other galaxies nearby which skew the results at small radii, but they 
remain consistent with a uniform surface density to within the uncertainties. 

\placefigure{RSD}

\subsection{What is the Redshift?}

While there is a significant excess of red galaxies in this field, their 
distribution on the sky does not support the interpretation that there 
is a cluster in the B3 0003$+$387 field.  The data are consistent with 
our looking through a wall or sheet of galaxies.  The obvious question 
is whether this structure is associated with the radio galaxy, or merely 
a line-of-sight superposition?  

A single color ($R-K^\prime$) is not sufficient for more sophisticated 
photometric redshift fitting, which normally requires photometry 
though three or more filters.  However, even passively-evolving 
elliptical galaxy colors do not become particularly red until the 
normally strong 4000\AA\ break is redshifted through the $R$ filter, 
which occurs at $z \simeq 0.8$ for the Harris $R$ filter used in this 
study.  Colors as red as ($R-K^\prime$) $>$ 5\fm2, corresponding to the {\em 
bluest} of our red excess galaxies, are difficult to explain unless 
the galaxies are at $z \gtrsim 1$.  The characteristics of the filters 
used in this project, and the resulting very red colors of the galaxies 
allow us to place a lower limit of $z \geq 0.8$ on their redshift.

We note that it is also possible to produce such red colors in a 
strong starburst at lower redshift, $z ~ 0.5$, assuming heavy extinction 
from dust.  However, the high surface density of red galaxies in 
this field would be very unusual, given the relatively low fraction 
of starbursts found in ERGs \citep{Cim99}.

Assuming the red excess galaxies are associated at a common redshift in 
some sort of wall or sheet structure, the luminosities of the brightest 
of our red excess galaxies can also place constraints on the redshift.  
We compare the brightest of our BRG subset, at $K^\prime \sim 17.7$, 
to the ``brightest cluster member'' galaxies seen in other high-redshift 
clusters.  These brightest cluster members follow a tight correlation 
in the $K$-band Hubble diagram \citep{Kz1,Kz2} up to at least a redshift 
of one.  Comparing our data to other high-redshift clusters 
(Table~\ref{BCGs}) shows that we are at least consistent with the 
$z > 1$ interpretation, and there is sufficient scatter in the $K$ 
magnitudes that our red excess galaxies could indeed be at $z = 1.47$.

\placetable{BCGs}

The origin of the activity in the central engines of radio galaxies can
often be traced to recent or ongoing mergers.  We note that there is a
bright and very red galaxy, with $K = 18.36$ and
($R-K$)\,$=$\,5\fm80 about 5 arcseconds ($\sim21\,h^{-1}$\,kpc) east of 
B3 0003$+$387.  In the $K$-band image from Keck (Figure~\ref{RKcomp}), 
this galaxy shows an asymmetric morphology, visible in the contours, 
with a spur or elongation in the direction of the radio galaxy.  There 
are additional faint features to the west of the radio galaxy and east 
of the red galaxy, roughly aligned with the axis between the two galaxies,
which may be the visible remnants of tidal tails created by
gravitational interaction.  However, these features are very faint
and may be attributable to image noise or superposition of another
object along the line of sight to the radio galaxy.  If this 
interaction is real, then this also supports the interpretation that 
the red galaxies in the B3 0003$+$387 field lie at $z = 1.47$, the 
redshift of the radio galaxy.

\section{Summary}

We have identified a significant overdensity of red galaxies in an 
extended field (44.3 square arcmin) around the $z = 1.47$ radio galaxy 
B3 0003$+$387.  This overdensity is as much as a factor of 14 higher 
than seen in equivalent blank-field surveys.  The spatial distribution 
of the galaxies is {\em not} consistent with their being associated in 
an evolved cluster centered on the radio galaxy, although we believe 
the evidence does strongly suggest that we are looking through a sheet 
or wall of galaxies associated with some extended Large-Scale Structure.  
The red colors of these galaxies are difficult to explain without the 
large cosmological K-corrections that would come with their being at 
high redshift ($z > 1$).  The presence of a powerful, high-redshift 
radio galaxy in the field, as well as its possible interaction with 
one of the red galaxies, suggest that this structure may lie at the 
same redshift as the radio source: $z = 1.47$.

It will be important to verify the reality of this high-redshift structure 
by obtaining spectroscopic redshifts for some fraction of the red 
galaxies in this field.  The brightest of the red excess galaxies are 
bright enough at visual wavelengths ($R$ $\sim$ 23-24$^{\rm m}$) to make 
spectroscopic followup possible with 8--10\,m-class telescopes.  In 
order to facilitate this followup, as well as further study of this 
field, we are making our full dataset available through the electronic 
version of the Astronomical Journal.

\acknowledgements

This project was partially supported by a Caltech Summer Undergraduate 
Research Fellowship (SURF) for OA.  We wish to thank S. Djorgovski for 
kindly providing the redshift for B3 0003$+$387, and G. Neugebauer for 
obtaining the Keck K-band image.  The NOAO WIYN-Queue staff did an 
excellent job of obtaining the $R$ data for this project.  We also 
thank Pat Hall for his helpful comments in refereeing this paper.


\newpage
\onecolumn

\clearpage
\pagestyle{empty}

\begin{deluxetable}{rrrrrrrrrrr}
   \footnotesize
   \tablewidth{0cm}
   \tablecaption{Photometric Catalog\tablenotemark{a} \label{catalog}}
   \tablehead{
      \colhead{x$_{K^\prime}$}		&
      \colhead{y$_{K^\prime}$}		&
      \colhead{$K^\prime$}		&
      \colhead{$\sigma_{K^\prime}$}	&
      \colhead{x$_R$}			&
      \colhead{y$_R$}			&
      \colhead{$R$}			&
      \colhead{$\sigma_R$}		&
      \colhead{$\alpha_{J2000}$}	&
      \colhead{$\delta_{J2000}$}	&
      \colhead{\#}			}
   \startdata
 2005.0 & 1775.7 & 15.97 & 0.03 & 2132.5 & 1663.5 & 18.63 & 0.03  & 00 06 07.55 & $+$39 02 09.6 & 1 \\
 1997.2 & 1348.7 & 19.51 & 0.14 & 2140.1 & 1234.6 & 22.12 & 0.05  & 00 06 07.56 & $+$39 00 44.9 & 2 \\  
 2008.1 & 2226.3 & 18.93 & 0.10 & 2119.4 & 2115.8 & 22.03 & 0.04  & 00 06 07.62 & $+$39 03 38.9 & 3 \\  
 1998.3 & 1751.7 & 18.91 & 0.10 & 2126.7 & 1639.2 & 24.52 & 0.28  & 00 06 07.66 & $+$39 02 04.8 & 4 \\  
 1990.5 & 1385.5 & 19.13 & 0.12 & 2132.1 & 1271.3 & 21.83 & 0.04  & 00 06 07.68 & $+$39 00 52.2 & 5 \\  
   \enddata
   \tablenotetext{a}{The complete version of this 
                 table is in the electronic edition of the Journal.  The 
                 printed edition contains only a sample.}
   \normalsize
\end{deluxetable}

\clearpage
\begin{deluxetable}{lrrr}
   \tablewidth{0cm}
   \tablecaption{Brightest Cluster Galaxies \label{BCGs}}
   \tablehead{
      \colhead{Cluster ID}	&
      \colhead{$z$}		&
      \colhead{$m_K$}		&
      \colhead{Ref}		}
   \startdata
      B3 0003$+$387    & 1.47?       & 17.7 & 1 \\
      ClG J0848$+$4453 & 1.27\phm{?} & 18.1 & 2 \\
      RXJ0848.9$+$4452 & 1.26\phm{?} & 16.7 & 3 \\
      AXJ2019$+$112    & 1.01\phm{?} & 17.1 & 4 \\
      Cl 1603$+$4329   & 0.92\phm{?} & 17.6 & 5 \\
      Cl 1603$+$4313   & 0.89\phm{?} & 17.0 & 5 \\
   \enddata
   \tablerefs{(1) This paper, $K^\prime$; (2) \citet{stanford}, $K$;  
              (3) \citet{Rosati}, $K_s$; (4) \citet{Benitez}, $K_s$; 
              (5) \citet{AS93}, $K$.}
\end{deluxetable}

\end{document}